\documentclass[10pt,conference]{IEEEtran}
\IEEEoverridecommandlockouts
\usepackage{cite}
\usepackage{amsmath,amssymb,amsfonts}
\usepackage{algorithmic}
\usepackage{graphicx}
\usepackage{textcomp}
\usepackage{xcolor}
\usepackage{subcaption}
\usepackage{booktabs} 
\usepackage{float}
\usepackage{hyperref}
\def\BibTeX{{\rm B\kern-.05em{\sc i\kern-.025em b}\kern-.08em
    T\kern-.1667em\lower.7ex\hbox{E}\kern-.125emX}}
\begin{document}

\title{Robust Bandwidth Estimation for Real-Time \\ \makebox[.9\linewidth]{Communication with Offline Reinforcement Learning}
}
\author{
    Jian Kai\textsuperscript{†}, Tianwei Zhang\textsuperscript{†}, Zihan Ling\textsuperscript{†}, Yang Cao\textsuperscript{†}, and Can Shen\textsuperscript{§} \\
    \textsuperscript{†}School of Electronic Information and Communications, Huazhong University of Science and Technology, China \\
    \textsuperscript{§}Zhongxing Telecommunication Equipment Corporation, China \\
    E-mail: \{kaji, twzhang, lingzh, ycao\}@hust.edu.cn, shen.can1@zte.com.cn
}
\maketitle
\begin{abstract}
Accurate bandwidth estimation (BWE) is critical for real-time communication (RTC) systems. Traditional heuristic approaches offer limited adaptability under dynamic networks, while online reinforcement learning (RL) suffers from high exploration costs and potential service disruptions. 
Offline RL, which leverages high-quality data collected from real-world environments, offers a promising alternative. However, challenges such as out-of-distribution (OOD) actions, policy extraction from behaviorally diverse datasets, and reliable deployment in production systems remain unsolved. 
We propose RBWE, a robust bandwidth estimation framework based on offline RL that integrates Q-ensemble (an ensemble of $Q$-functions) with a Gaussian mixture policy to mitigate OOD risks and enhance policy learning. A fallback mechanism ensures deployment stability by switching to heuristic methods under high uncertainty. 
Experimental results show that RBWE reduces overestimation errors by 18\% and improves the 10th percentile Quality of Experience (QoE) by 18.6\%, demonstrating its practical effectiveness in real-world RTC applications. The implementation is publicly available at \url{https://github.com/jiu2021/RBWE_offline}.
\end{abstract}


\section{Introduction}  
In recent years, applications such as video conferencing, remote education, and cloud gaming leveraging real-time communication (RTC) systems have become increasingly popular. Both academia and industry have been actively investigating approaches to enhance the quality of experience (QoE, user's subjective experience of audio and video quality, latency, and other factors) in real-time communication systems.

In RTC systems, the bandwidth estimation (BWE) serves as the target bitrate for audio/video encoders and controls the client’s sending rate. The system must accurately estimate the dynamically changing link capacity between the sender and receiver to ensure a seamless communication experience. Mainstream BWE methods can be broadly classified into two categories: rule-based heuristics and data-driven approaches. WebRTC \cite{alvestrand2021rfc}, as a concrete implementation of an RTC system, employs the Google Congestion Control (GCC) algorithm \cite{carlucci2016analysis}, which serves as a representative heuristic approach. GCC estimates the current bandwidth by analyzing delay jitter and packet loss. However, it fails to generalize effectively in complex and dynamic real-world network environments \cite{xiao2023ember}. Data-driven methods typically formulate the BWE problem as a Markov decision process (MDP) and leverage reinforcement learning (RL) techniques to derive optimal decision policies. OnRL \cite{zhang2020onrl} trains models through online RL in a production environment, which~is~costly~and~involves~exploration~behaviors~prior~to~model~convergence,~potentially~leading ~to~unpredictable QoE degradation. Although some approaches leverage simulators for training, they still suffer from the sim-to-real gap, reducing effectiveness in real-world deployment \cite{eo2022opennetlab, khairy2024acm}.

Recent research increasingly explores offline RL to tackle these problems~\cite{tan2024accurate, lu2024pioneer}. Offline RL does not require interaction with the environment, thereby mitigating the risk of catastrophic failures during model optimization\cite{prudencio2023survey}. Through extensive observation and analysis of data produced by diverse BWE policies (commonly known as behavior policies), offline RL holds the potential to discover the optimal~policy. However, several challenges remain:  
(1) The algorithm must balance minimizing deviation from behavior policies with optimizing policy learning to prevent catastrophic out-of-distribution (OOD) actions.  
(2) How can the algorithm effectively extract the best policy from heterogeneous behavior policies incorporated in the dataset?  
(3) How can a model trained solely on an offline dataset be reliably transferred to a real-world online deployment?

This paper presents RBWE, a robust bandwidth estimation framework developed to tackle the challenges above. The proposed framework consists of two stages: offline training and online deployment, ensuring high-precision bandwidth estimation while preserving system security and stability.  
\textbf{Offline Training Stage:} RBWE leverages offline RL to learn an optimal BWE policy from large-scale RTC session data. A key innovation is the integration of Q-ensemble (an ensemble of $Q$-functions) with a Gaussian mixture policy to precisely estimate state-action values and capture the multi-modal characteristics of diverse behavior policies.  
\textbf{Online Deployment Stage:} During inference, RBWE leverages Q-ensemble to quantify the uncertainty of candidate actions. When the model exhibits sufficient confidence, RBWE extracts conservative actions from the Gaussian mixture policy via numerical optimization to enable robust bandwidth estimation. Otherwise, the system falls back to a rule-based algorithm to maintain stability.

We assessed RBWE's performance in both offline and online environments. In the offline evaluation, RBWE reduced the overestimation error rate by 18\% compared to the behavior policy, which is a critical prerequisite for RTC system stability. The online evaluation was performed on our custom-built RTC experimental platform. RBWE achieved the highest QoE, outperforming GCC by 18.6\% in the 10th percentile QoE.

Our main contributions are as follows:
\begin{itemize}
    \item We utilize Q-ensemble to mitigate OOD issues in offline training and adopt Gaussian mixture models to effectively capture heterogeneous behavior policies.
    \item We construct a controlled RTC online experimental platform and introduce a robust fallback mechanism based on model output uncertainty.
    \item Experimental results show that RBWE outperforms both GCC and the latest offline RL method in QoE, notably reducing low-QoE occurrences.
\end{itemize}

The remainder of this paper is structured as follows. Section \ref{relatedwork} reviews related work. Section \ref{Problem_Formulation} formulates the problem of bandwidth estimation in RTC. Section \ref{method11} details our approach in two stages. Section \ref{evaluation} presents the specific implementation of our method and the evaluation. Finally, Section \ref{conclusion} summarizes our findings and conclusions.
\section{Related Work} \label{relatedwork}
\textbf{Rule-based heuristics.} Today's video conferencing applications (such as Microsoft Teams, Google Hangouts, and Tencent Meeting) all use rule-based BWE algorithms and are usually redesigned with GCC-inspired mechanisms \cite{agarwal2025mowglipassivelylearnedrate}. This allows them to design different rules for different scenarios to improve QoE stability for users, ensuring robustness and stability in production environments.

\textbf{Data-driven approaches.} In contrast, researchers have explored using data-driven approaches (especially RL) to generate alternative rate control algorithms. OnRL and Concerto \cite{zhou2019learning} build non-standalone estimators directly on top of heuristic methods or utilize heuristics as fall-back mechanisms to mitigate unreliable agent behavior in tail scenarios. Loki \cite{zhang2021loki} explores fusing cloned heuristic features with online RL-based models for video bitrate prediction. 

Microsoft Research proposed training bandwidth estimation models through offline RL and have demonstrated the potential of offline RL to enhance the QoE for users in RTC \cite{khairy2024acm}. By leveraging a diverse dataset derived from Microsoft Teams calls and incorporating objective audio/video quality scores, which have a high correlation with subjective experience scores as rewards, they have paved the way for the development of more user-centric BWE models. Building upon these insights, previous works \cite{tan2024accurate, lu2024pioneer} have primarily concentrated on neural architecture design or offline RL algorithm selection. However, most overlook the heterogeneity of behavior policies and deployment challenges posed by real-world distribution shifts. To bridge this gap, we propose RBWE, a policy extraction framework coupled with a robust deployment mechanism tailored for RTC systems.

\section{Problem Formulation} \label{Problem_Formulation}
We formulate the bandwidth estimation problem in RTC as a MDP defined by the tuple $\langle \mathcal{S}, \mathcal{A}, P, r, \rho_0, \gamma \rangle$, where $\mathcal{S}$ denotes the state space consisting of network and media features, and $\mathcal{A}$ is the action space representing bandwidth decisions. The transition dynamics $P(s'|s, a)$ model the evolution of network conditions in response to actions, and the reward function $r(s, a)$ reflects the communication quality. $\rho_0$ is the initial state distribution, and $\gamma \in [0, 1]$ is the discount factor. In the offline RL setting, the agent learns from a fixed dataset $\mathcal{D} = \{(s, a, r, s')\}$ collected under various BWE policies in production environments. Without further environment interaction, the objective is to learn a policy $\pi(a|s)$ that maximizes the expected discounted return estimated from $\mathcal{D}$:
\begin{equation}
\max_\pi J_{\mathcal{D}}(\pi) = \mathbb{E}_{(s_t, a_t, r_t, s_{t+1}) \sim \mathcal{D}} \left[ \sum_{t=0}^{\infty} \gamma^t r(s_t, a_t) \right].
\end{equation}
Dataset $\mathcal{D}$ is collected from diverse behavior policies under varying network conditions, resulting in significant heterogeneity and distributional shift. Additionally, network dynamics are often non-stationary, and feedback signals may be delayed or noisy, complicating the estimation of reliable state-action value functions ($Q$-functions). These characteristics challenge standard offline RL methods and motivate the need for tailored solutions for robust policy extraction and deployment in RTC systems.

\section{Method} \label{method11}
\begin{figure}[b]
    \vspace{-1.5em}
    \centering
    \includegraphics[width=0.45\textwidth]{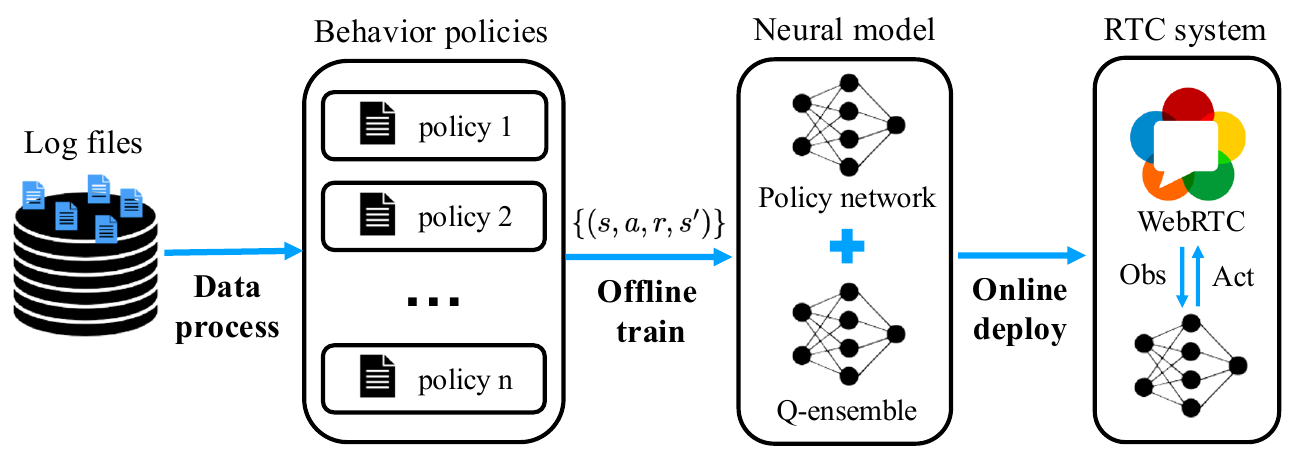}
    \caption{RBWE overview.}
    \label{method}
\end{figure}

Motivated by the challenges, we present RBWE, a data-driven bandwidth estimation framework. As illustrated in Figure~\ref{method}, RBWE comprises two key stages: offline training and online deployment. Section \ref{Method-B} provides a detailed description of the offline RL algorithm and its integration with the Gaussian mixture policy and Q-ensemble. Section \ref{Method-C} elaborates on the online deployment strategy, which consists of two key components: conservative action estimation and OOD detection.
\subsection{Offline Training}\label{Method-B}
\paragraph{\textbf{State, Action and Reward}} We utilize public data from Microsoft Teams to construct the dataset $\mathcal{D} = \{(s, a, r, s')\}$, where states encode network statistics, actions represent bandwidth decisions, and rewards reflect QoE-related metrics.
\begin{itemize}
\item State: The state is a 150-dimensional vector, constructed from 15 types of network statistics collected across 5 short-term and 5 long-term monitoring intervals. Example features include receiving rate, queuing delay, packet loss ratio, and video packet probability.  A full list of features is provided in \cite{khairy2024acm}.
\item Action: We define the action as the ratio of the original estimated bandwidth $E$ and the received rate $R$, and normalize it using a logarithmic mapping: $a=log(\frac{E}{R})$. 
\item Reward: To achieve a QoE-driven learning objective, we formulate the reward function as $r(s,a) = (2 - \alpha) \cdot q_a + \alpha \cdot q_v$, using dataset-provided mean opinion score for audio ($q_a$) and video ($q_v$). Given the video quality is more sensitive to network changes, we set $\alpha = 1.8$.
\end{itemize}

\paragraph{\textbf{Learning algorithm}} 
Learning a policy that outperforms the behavior policy requires evaluating actions absent from the dataset. This necessitates balancing policy improvement against distributional shift, as actions far from the data distribution are harder to estimate accurately. To address this, RBWE leverages implicit Q-learning (IQL)~\cite{kostrikov2021offlinereinforcementlearningimplicit} for in-sample learning, enabling more reliable $Q$-function estimation and policy improvement through a two-stage process.

In the policy evaluation stage, IQL uses the expectile regression to approximate the optimal value function \(V(s)\): 
\begin{equation}
\mathcal{L}_V(\psi)=\mathbb{E}_{(s,a)\sim\mathcal{D}}\left[L_2^\tau\left(Q_{\hat{\theta}}(s,a)-V_{\psi}(s)\right)\right], \label{value_loss}
\end{equation}
where \(L_2^\tau(u)=|\tau-\mathbf{1}(u<0)|u^2\) is an asymmetric \(l_2\) loss, and \(Q_{\hat{\theta}}(s,a)\) is the target state-action network. The state-action value function \(Q_{\theta}(s, a)\) is updated by minimizing the temporal difference loss: 
\begin{equation}
\mathcal{L}_Q(\theta)=\mathbb{E}_{(s,a,s')\sim\mathcal{D}}\bigl[(r(s,a)+\gamma V_\psi(s')-Q_\theta(s,a))^2\bigr]. \label{q_loss}
\end{equation}

In the policy extraction stage, IQL uses advantage-weighted regression (AWR) by  minimizing the loss for optimizing the final policy \(\pi_{\phi}(s)\) is:
\begin{equation}
\mathcal{L}_\pi(\phi)=\mathbb{E}_{(s,a)\sim\mathcal{D}}[\exp(\beta\cdot A(s,a))\log\pi_\phi(a|s)], \label{policy_loss}
\end{equation}
where the advantage \(\begin{aligned} A(s,a)=Q_\theta(s,a)-V_\psi(s) \end{aligned}\), and  \(\beta\in[0,\infty)\) is used to adjust the balance between maximizing Q-values and behavior cloning. Note that these losses do not use any explicit policy and only utilize actions from the dataset for both objectives. This makes it easier to model the policy output directly as a distribution rather than a specific estimated bandwidth value, as shown in the following text.

\paragraph{\textbf{Q-ensemble}} 
Due to the policy extraction method based on AWR, the learning of the \(Q\)-function directly determines the quality of the final policy. A common consensus in offline RL is the tendency to underestimate Q-values, and assigning a lower Q-value also plays a crucial role in BWE tasks: (1) Suppressing model overestimation: A higher bandwidth estimate improves audio-video quality, leading to a higher corresponding Q-value. However, even slight overestimation beyond available bandwidth can cause severe congestion, degrading users' QoE. Therefore, we need to suppress the model's overestimation by underestimating the Q value.
(2) Penalizing corrupted data: The presence of extreme data points, which can be considered OOD samples, leads to a pronounced heavy-tail effect in policy outputs. To improve model robustness, it is necessary to apply underestimation in such cases.

IQL uses a version of clipped double Q-learning (CDQ), taking a minimum of two \(Q\)-functions for \(V\)-function and policy update, which relates to methods that consider the confidence bound of the Q-value estimates. Q-ensemble explicitly expresses the uncertainty-based penalization term in CDQ, and is achieved by simply increasing the number of Q-networks \cite{an2021uncertainty}. Suppose $Q(s, a)$ follows a Gaussian distribution with mean $m_q(s, a)$ and standard deviation $\sigma_q(s, a)$. Also, let $\{{Q}_j(s,a)\}_{j=1}^N$ be realizations of $Q(s, a)$. Then, we can approximate the expected minimum of the realizations following the work of \cite{royston1982expected} as:
\vspace{-0.2em}
{\small \begin{equation} \mathbb{E}[\min_{j=1,\dots,N}\!Q_j(s,a)]\!\approx\!m_q(s,a)\!-\!\Phi^{-1}\!\left(\frac{N-\frac{\pi}{8}}{N-\frac{\pi}{4}+1}\right)\sigma_q(s,a), \label{min_q} \end{equation} }\vspace{-0.2em}where $\Phi$ is the cumulative distribution function (CDF) of the standard Gaussian distribution. This relation indicates that using the clipped Q-value is similar to penalizing the ensemble mean of the Q-values with the standard deviation scaled by a coefficient dependent on $N$. 

The value function \( V_{\psi}(s) \) is implemented as a two-layer fully connected (FC) neural network. The Q-function \( Q_\theta(s,a) \) is modeled as an ensemble of \( N \) Q-networks, each sharing the same architecture as the value function. To mitigate the impact of heavy-tailed distributions, we replace the standard squared loss in~\eqref{q_loss} with the Huber loss in each Q-network update.

\begin{figure}[bp]
    \vspace{-1.5em}
    \centering
    \includegraphics[width=0.4\textwidth]{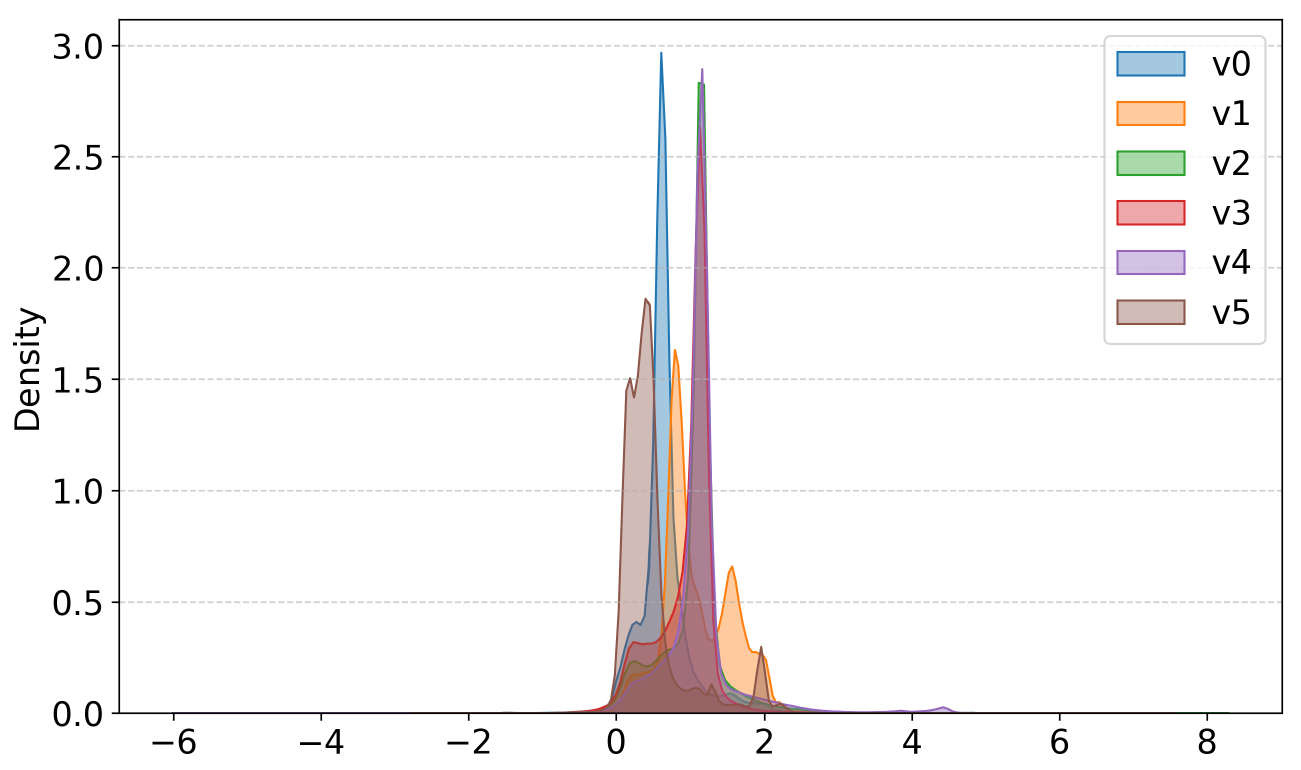}
    \caption{Output probability distribution of behavior policies.}
    \label{fig_actions_kde}
\end{figure}

\paragraph{\textbf{Gaussian mixture policy}} There are substantial differences in the performance of various behavior policies. 
To analyze heterogeneous behavior policies, we applied kernel density estimation to the probability distributions of their outputs. As illustrated in Fig.~\ref{fig_actions_kde}, although the actions have been normalized, the distributions of different policies vary considerably, and the action distribution of a single policy does not follow a unimodal Gaussian distribution. This analysis supports our choice of employing a Gaussian mixture model for a more accurate representation of behavior policies.
Traditional Gaussian policy networks commonly assume that the probability distribution of an action $a$ given a state $s$ follows a single Gaussian distribution:
\vspace{-0.2em}
\begin{equation}
\pi(a|s)=\mathcal{N}(a;\mu(s),\sigma(s)),
\end{equation}
\vspace{-0.2em}where $\mathcal{N}$ denotes the Gaussian distribution function, $\mu(s)$ and $\sigma(s)$ are output by the policy network as the mean and standard deviation of the actions, respectively. To avoid the mode collapse issue caused by a unimodal Gaussian distribution, we extend it to a weighted mixture of $K$ Gaussians:
\vspace{-0.2em}
\begin{equation}
\pi(a|s)=\sum_{k=1}^K p_k(s)\,\mathcal{N}\!\left(a;\mu_k(s),\sigma_k(s)\right)
\end{equation}
\vspace{0.2em}where $p_k(s)$ are the mixing weights with $\sum_{k=1}^K p_k(s) = 1$.

\begin{figure}[bp]
    \vspace{-1.5em}
    \centering
    \includegraphics[width=0.3\textwidth]{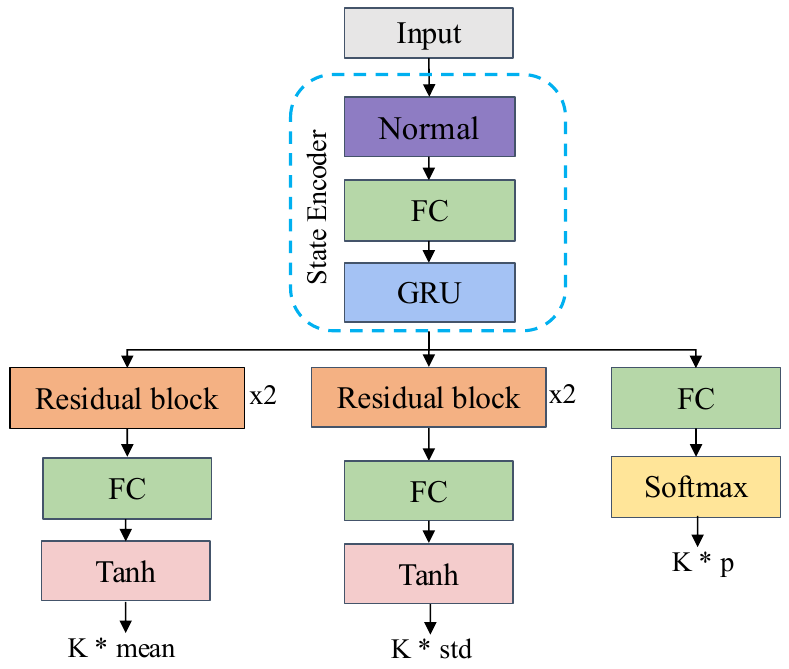}
    \caption{Network architecture of Gaussian mixture policy.}
    \label{policy_network}
\end{figure}

The implementation of the Gaussian mixture policy is shown in Figure~\ref{policy_network}. The network initially applies a normalization layer, followed by an FC layer and a gated recurrent unit, to encode shared state representations. It then splits into three branches, each responsible for parameterizing \( \mu_k(s) \), \( \sigma_k(s) \), and \( p_k(s) \). Mean branch: Two residual modules, each containing two FC layers with activation functions, followed by a final FC layer with a Tanh activation; Standard deviation branch: Identical to the mean branch but with independent parameters; Mixture weight branch: Uses an FC layer and Softmax activation to ensure weights sum to one. The network parameters are updated using \eqref{policy_loss}. 

\subsection{Online Deploy}\label{Method-C}
\paragraph{\textbf{Conservative Action Estimation}}  
For a given state $s$, the Gaussian mixture policy outputs distribution parameters: \( \mu_k(s) \), \( \sigma_k(s) \), and \( p_k(s) \). Our goal is to find the action \( a^* \) that maximizes the mixture density \( \pi(a|s) \). As the maximization lacks a closed-form solution due to the multimodal distribution, we equivalently reformulate it as minimizing the negative log-likelihood, exploiting the monotonicity of the logarithm:
\begin{equation}
a^* = \arg\max_a \pi(a|s)\!~\Leftrightarrow\!~a^* = \arg\min_a [-\log \pi(a|s)].
\end{equation}
This optimization is solved numerically using the Limited-memory Broyden–Fletcher–Goldfarb–Shanno algorithm with a predefined precision threshold. To characterize the local behavior of the distribution around \( a^* \), we perform a second-order Taylor expansion: 
\vspace{-0.2em}
\begin{equation}
\log \pi(a|s) \approx \log \pi(a^*|s) - \frac{1}{2} \lambda^* (a - a^*)^2,
\end{equation}
\vspace{-0.2em}where $\lambda^* = -\frac{d^2}{da^2} \log \pi(a|s) \Big|_{a=a^*}$. Based on this, we define the effective local standard deviation as $\sigma^* = \frac{1}{\lambda^*}$.

Finally, by combining \( a^* \) and \( \sigma^* \), we derive the lower-confidence bound (LCB) for conservative action selection:  
\begin{equation}
a_{\text{chosen}} = a^* - \delta \cdot \sigma^*, \label{eq19}
\end{equation}
where \( \delta > 0 \) is a risk control parameter. When \( \sigma^* \) is large, the LCB mechanism biases the action selection toward a more conservative estimate, mitigating the risk of overestimation. Conversely, when \( \sigma^* \) is small, the model tends to choose \( a^* \), indicating a high confidence level in the decision.

\paragraph{\textbf{OOD Detection}} 
Naive offline RL policy struggles to address OOD transition dynamics arising from exogenous processes affecting the environment post-deployment. We determine whether to adopt the action proposed by the current policy by leveraging Q-ensemble guidance during inference. Let $\{Q_{\theta_i}\}_{i=1}^{N}$ denote the Q-value estimates for the current state-action pair $(s, a)$ obtained from $N$ Q-networks. The mean and standard deviation of these estimates are $m_q(s,a)$ and $\sigma_q(s,a)$, respectively. To obtain a scale-invariant measure, we define the relative uncertainty as: 
\vspace{-0.2em}
\begin{equation}
U_q(s,a) = \frac{\sigma_q(s,a)}{\max(|m_q(s,a)|, \epsilon)}, \label{eq20}
\end{equation}
\vspace{-0.2em}where $\epsilon$ is a small constant to prevent numerical instability when the mean approaches zero. We introduce an uncertainty threshold $\tau_u$ to evaluate $U_q(s,a)$; if $U_q(s,a) > \tau_u$, it indicates high ensemble disagreement caused by two primary factors: (1) the model encountering an unfamiliar state or (2) the model failing to learn a stable action. In such cases, the policy estimates become unreliable, necessitating a fallback to the original GCC algorithm for bitrate control.

\begin{table*}[ht]
    \centering
    \vspace{0.2em}
    \begin{minipage}{0.34\textwidth} 
        \begin{center}
        \caption{Estimated accuracy}
        \label{tab:offline_results}
        \renewcommand{\arraystretch}{1.2}
        \setlength{\tabcolsep}{6pt}
        \begin{tabular}{lcccc}
            \toprule
            Measurement & $mse$ & $e$ & $e^+$ & $e^-$ \\
            \midrule
            RBWE & 3.49 & 0.27 & 0.31 & 0.24 \\
            behavior policies & 3.36 & 0.25 & 0.49 & 0.20  \\
            Schaferct & 2.64 & 0.30 & 0.48 & 0.17 \\
            \bottomrule
        \end{tabular}
        \end{center}
    \end{minipage}
    \hfill
    \begin{minipage}{0.64\textwidth} 
        \begin{center}
        \caption{Scores of rate, delay, loss, network, video, and QoE (Mean ± Std)}
        \label{tab:online_results}
        \renewcommand{\arraystretch}{1.2}
        \setlength{\tabcolsep}{6pt}
        \begin{tabular}{lcccccc}
            \toprule
            Measurement & $S_{rate}$ & $S_{delay}$ & $S_{loss}$ & $S_{network}$ & $S_{video}$ & QoE \\
            \midrule
            RBWE & 71.0±11.1 & 43.8±24.3 & 99.6±0.9 & 52.8±4.0 & 82.5±10.4 & 67.6±5.6 \\
            Schaferct & 66.4±10.6 & 41.8±23.4 & 98.7±5.7 & 51.2±5.3 & 80.6±10.1 & 66.0±5.5 \\
            GCC & 57.3±15.1 & 61.2±16.7 & 100.0±0.3 & 53.7±4.5 & 78.3±17.4 & 65.9±10.3 \\
            \bottomrule
        \end{tabular}
        \end{center}
    \end{minipage}
\end{table*}
\begin{figure*}[ht]
    \centering
    \begin{subfigure}{0.24\textwidth}
        \centering
        \includegraphics[width=\linewidth]{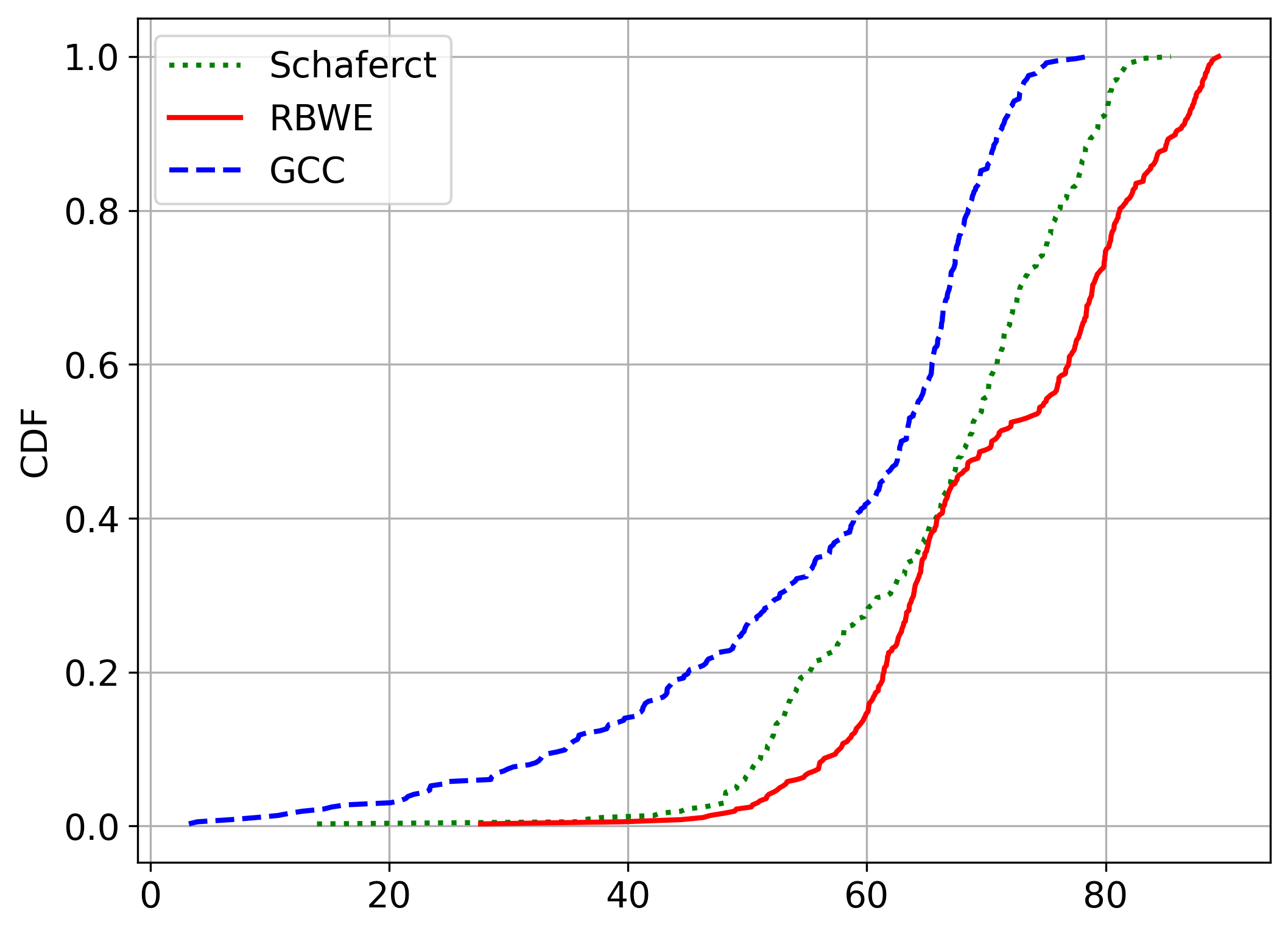}
        \vspace{-1em}
        \caption{Bandwidth Utilization}
        \label{fig:a}
    \end{subfigure}
    \hfill
    \begin{subfigure}{0.24\textwidth}
        \centering
        \includegraphics[width=\linewidth]{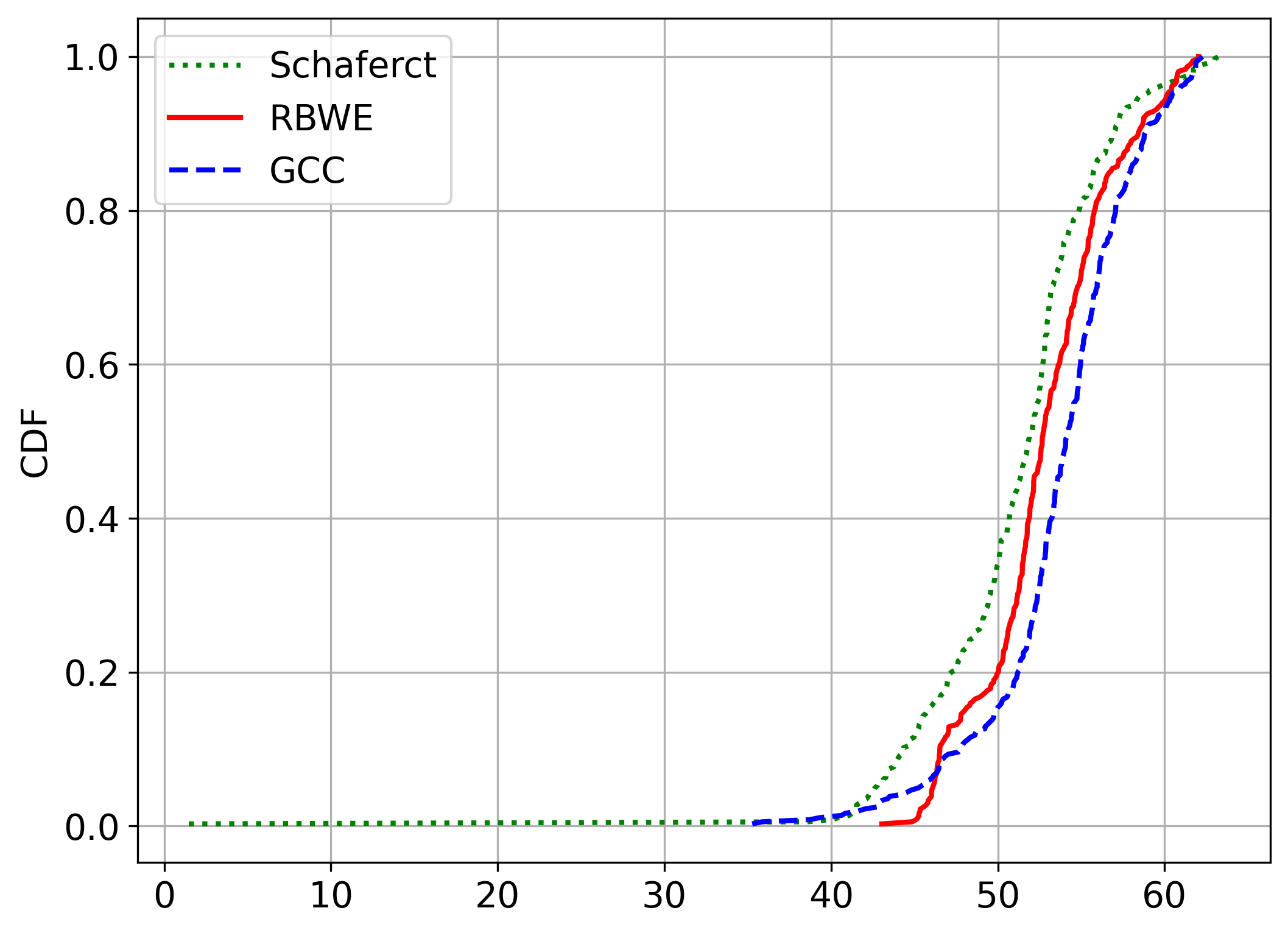}
        \vspace{-1em}
        \caption{Network Score}
        \label{fig:b}
    \end{subfigure}
    \hfill
    \begin{subfigure}{0.24\textwidth}
        \centering
        \includegraphics[width=\linewidth]{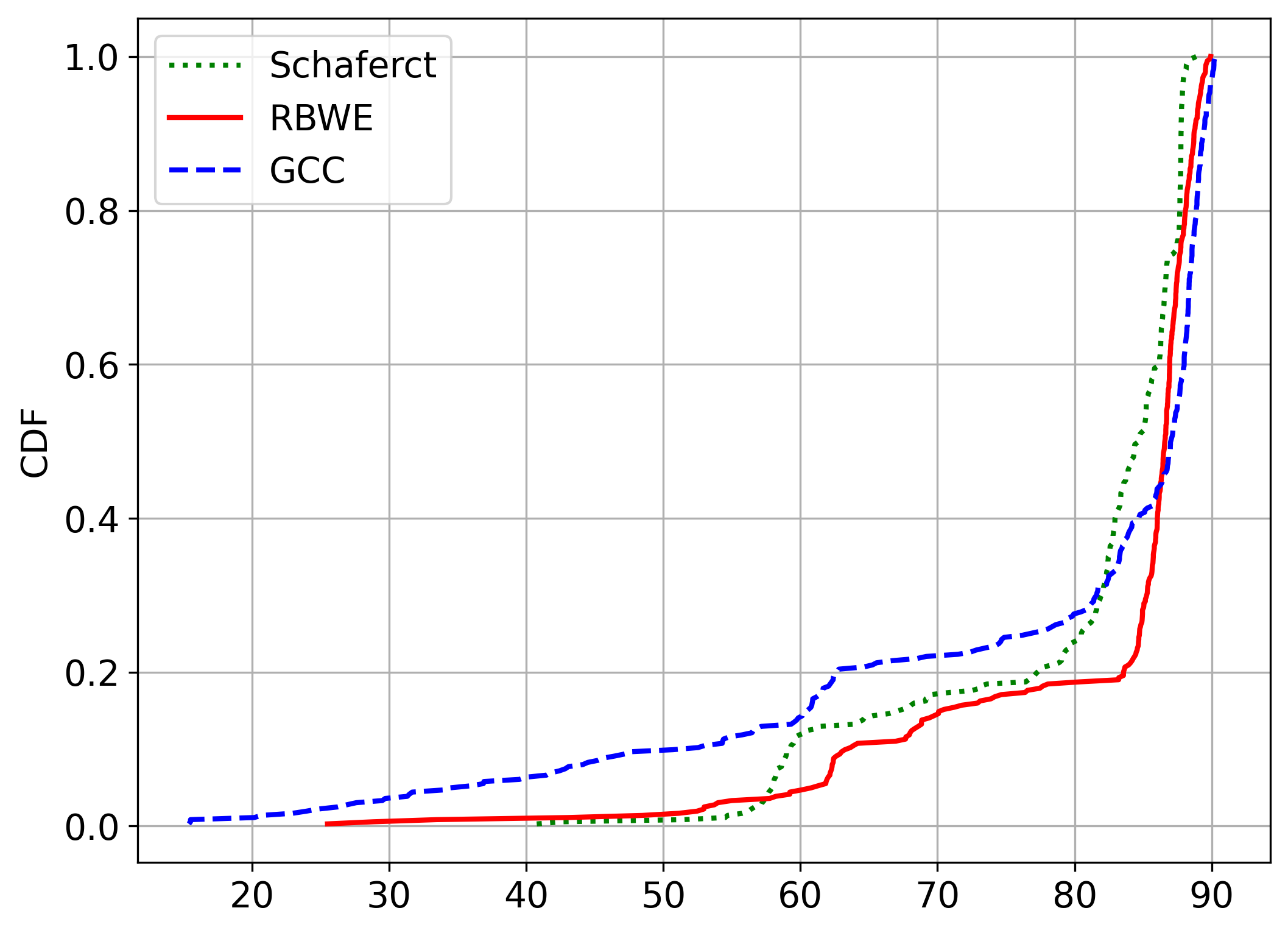}
        \vspace{-1em}
        \caption{Video Quality}
        \label{fig:b}
    \end{subfigure}
    \hfill
    \begin{subfigure}{0.24\textwidth}
        \centering
        \includegraphics[width=\linewidth]{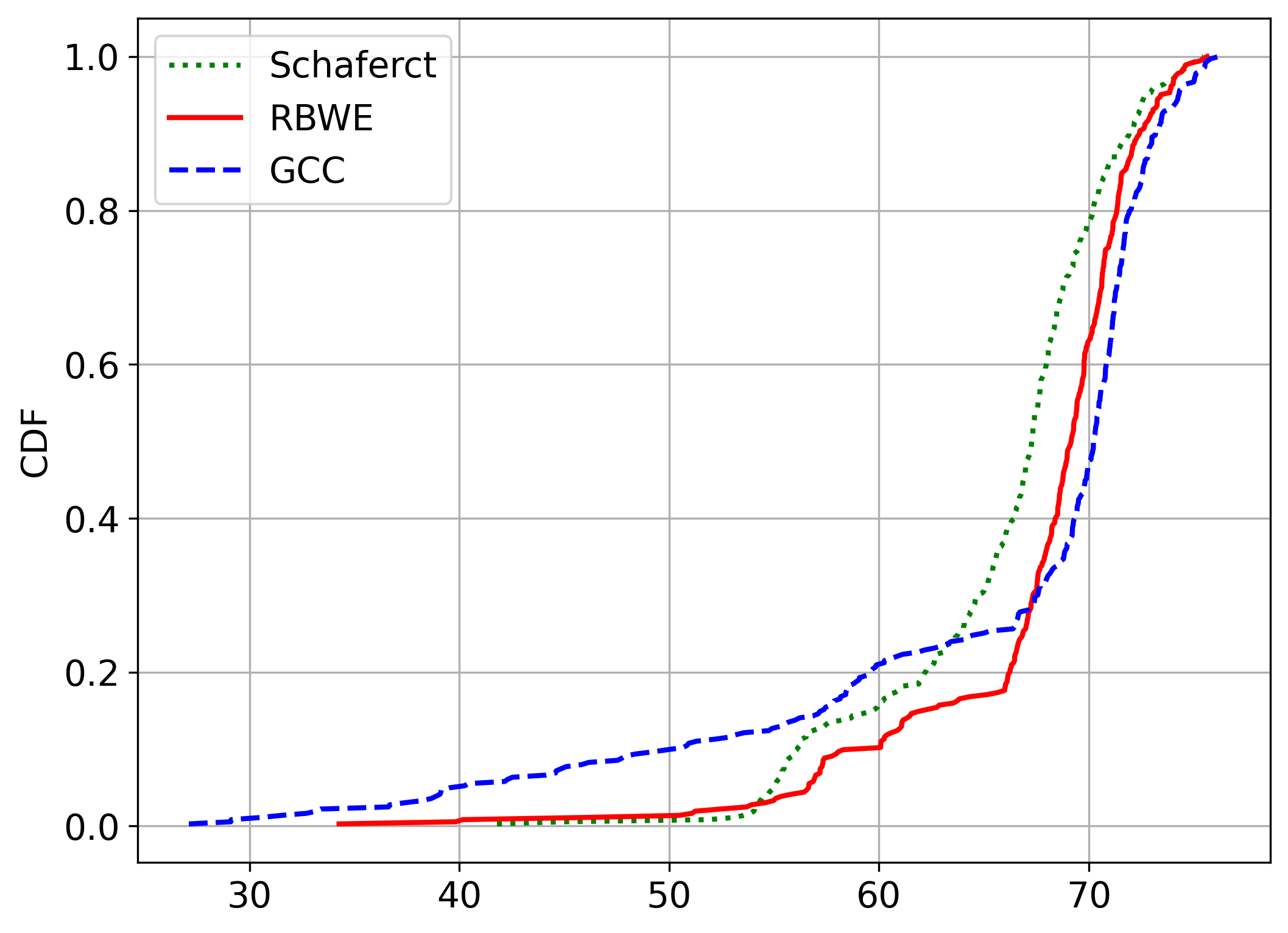}
        \vspace{-1em}
        \caption{QoE}
        \label{fig:d}
    \end{subfigure}

    \caption{CDFs of Bandwidth Utilization, Network Score, Video Quality, and QoE.}
    \label{fig:four_images_cdf}
\end{figure*}
\begin{figure*}[ht]
\vspace{-0.5em}
    \centering
    \begin{subfigure}{0.24\textwidth}
        \centering
        \includegraphics[width=\linewidth]{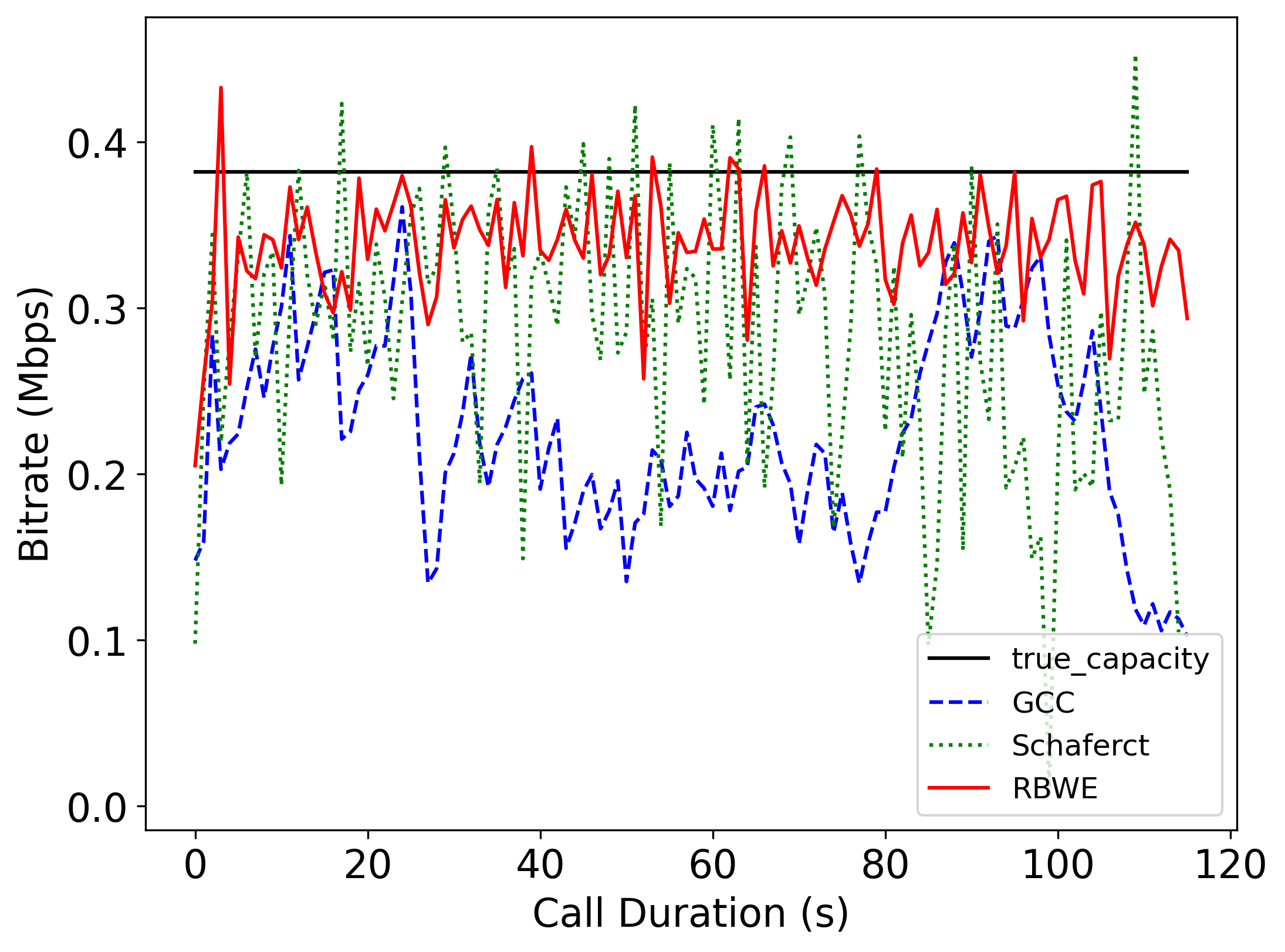}
    \end{subfigure}
    \hfill
    \begin{subfigure}{0.24\textwidth}
        \centering
        \includegraphics[width=\linewidth]{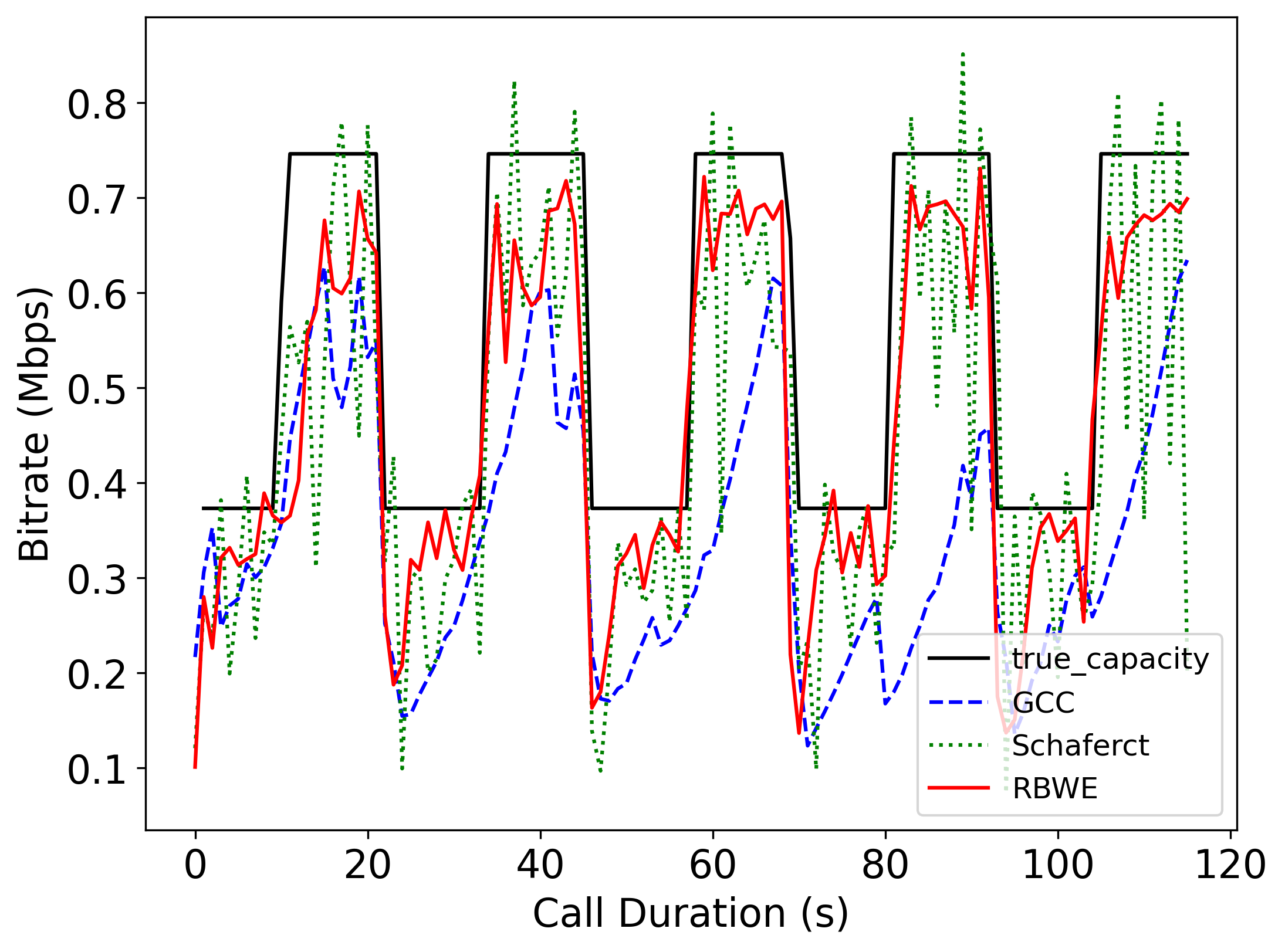}
    \end{subfigure}
    \hfill
    \begin{subfigure}{0.24\textwidth}
        \centering
        \includegraphics[width=\linewidth]{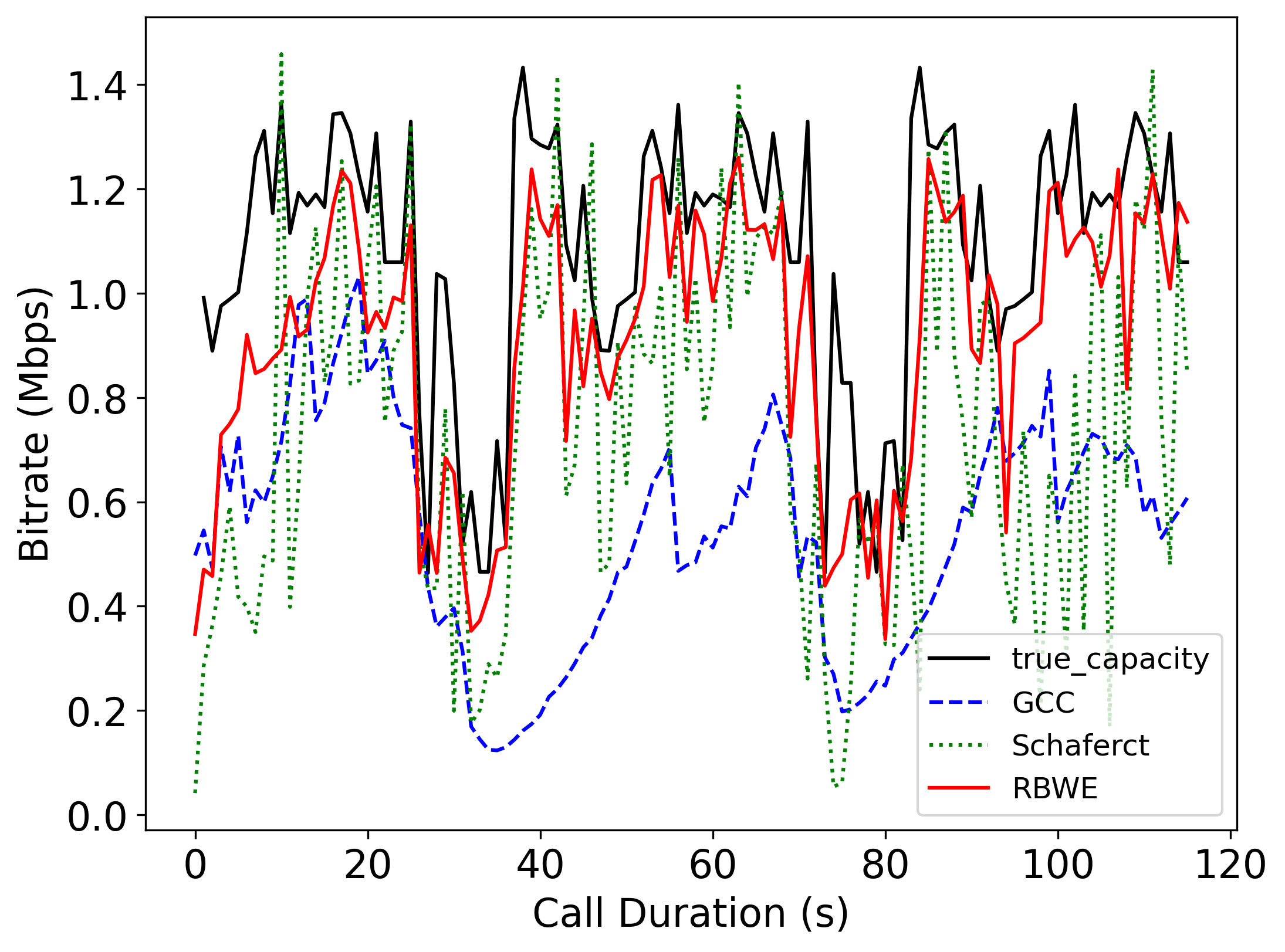}
    \end{subfigure}
    \hfill
    \begin{subfigure}{0.24\textwidth}
        \centering
        \includegraphics[width=\linewidth]{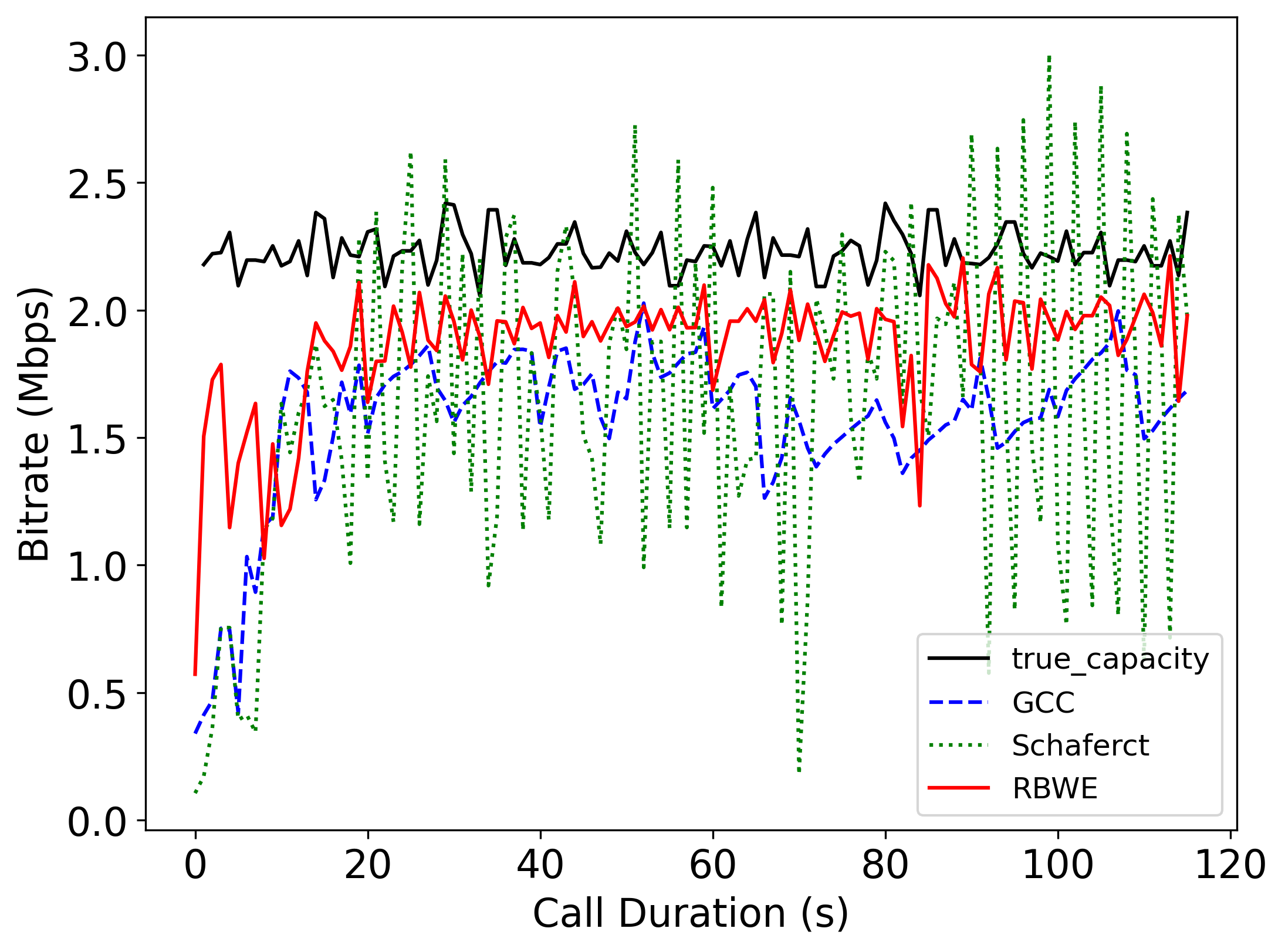}
    \end{subfigure}
    \caption{Bandwidth estimation under four dynamic network traces.}
    \label{fig:four_images_trace}
    \vspace{-1em}
\end{figure*}

\section{Evaluation} \label{evaluation} 
\subsection{Experimental Setup}
\paragraph{\textbf{Dataset}} 
Microsoft offers a comprehensive dataset comprising data from 18,859 real-world calls and 9,405 emulated test calls \cite{khairy2024acm}. We randomly selected 1,800 calls from each behavior policy and extracted millions of state transitions \((s, a, r, s^{\prime})\) for model training. For evaluation, we replay the network traces from the emulated calls.

\paragraph{\textbf{Offline settings}} 
The Q-ensemble comprises \(N=10 \) Q-networks, and the mixture policy contains \(K=4 \) Gaussian components. All neural networks consist of 256 hidden units. The hyperparameter settings for the IQL algorithm are \( \tau = 0.7 \) in \eqref{value_loss}, \( \gamma = 0.99 \) in \eqref{q_loss}, and \( \beta = 3 \) in \eqref{policy_loss}. The model is implemented in PyTorch and trained on an NVIDIA GeForce RTX 4090 with a batch size of 512 and a learning rate of \( 3 \times 10^{-4} \) for all networks.

\paragraph{\textbf{Online system}} 
We developed a complete RTC system for online evaluation. The agent is implemented in Python and interacts with WebRTC's C++ core. Frame- and packet-level events are transmitted to the agent via sockets, while the agent's decisions are written to shared memory and subsequently accessed by WebRTC for media flow control. At the sender side, state information is computed every 60~ms based on RTCP feedback and passed to the policy network for bandwidth estimation. The resulting action, along with the corresponding state, is forwarded to the Q-ensemble for OOD action detection. Empirically, we set $\delta = 0.5$ in \eqref{eq19} and $\tau_u = 0.4$ to balance detection sensitivity and performance stability. If the action passes OOD detection, it replaces the target bitrate computed by GCC; otherwise, the original GCC bitrate is retained. During deployment, all networks are converted to the ONNX format (30~MB) and achieve an inference speed of about 1~ms on our test platform (Ubuntu 22.04 with an AMD EPYC 7302 processor), and the system can also be deployed on resource-constrained edge devices.

\paragraph{\textbf{Benchmarks}} 
For fair comparison, we exclude online RL algorithms such as OnRL, which are challenging to reproduce faithfully due to differences in environment settings, and instead evaluate against the following baselines:
\begin{itemize}
    \item Schaferct \cite{tan2024accurate}: The current state-of-the-art (SOTA) method, also based on offline RL. We directly evaluate its publicly available ONNX model.
    \item Behavior policies: Heuristic or model-based strategies provided in the dataset, derived from UKF, GCC, and other data-driven algorithms.
    \item GCC: The default WebRTC congestion control algorithm, which follows an additive-increase/multiplicative-decrease (AIMD) strategy to adjust the sending bitrate.
\end{itemize}

\subsection{Offline Evaluation}
The \textbf{offline evaluation} process involves directly feeding the observed states from diverse behavior policies into our model. The primary objective is to assess whether our policy can surpass the limitations of the behavior policies and produce more accurate predictions.

\textbf{Metrics.} To quantify estimation accuracy, we evaluate model performance using three key metrics as defined in \cite{khairy2024acm}: mean squared error ($mse$), overall error rate ($e$), overestimation error rate ($e^+$), and underestimation error rate ($e^-$).

We tested all models on the 9,405 emulated test calls. The results are presented in Table~\ref{tab:offline_results}, showing the mean of each metric. As previously discussed, mitigating substantial overestimation is essential for ensuring safe online deployment. Our model substantially decreases the likelihood of overestimation while preserving both the overall error rate and the underestimation rate. Specifically, compared to Schaferct and Behavior policies, the overestimation rate is reduced by 17\% and 18\%, respectively. It is important to note that $mse$ is an absolute metric and can be significantly affected by traces with high bandwidth. In this case, a slight underestimation can cause $mse$ to soar.

\subsection{Online Evaluation}
The \textbf{online evaluation} deploys the model in the RTC system to primarily evaluate whether our approach can effectively enhance the user's QoE. 
\begin{equation}
\begin{aligned}
S_{rate} & =100\times U \\
S_{delay} & =100\times\frac{d_{\max}-d_{95\mathrm{th}}}{d_{\max}-d_{\min}} \\
S_{loss} & =100\times(1-L) \\
S_{network} & =0.2S_\mathrm{delay}+0.3S_{\mathrm{loss}}+0.2S_{\mathrm{rate}} \\
S_{video} & =\mathrm{VMAF} \\
QoE &=0.5S_\mathrm{network}+0.5S_\mathrm{video}
\end{aligned} \label{net_score}
\end{equation}

\textbf{Metrics.} The design of the overall QoE is formulated in \eqref{net_score} and already widely used \cite{eo2022opennetlab}. The overall QoE is the average of the video score and the network score. The network score is a weighted combination of throughput, packet delay, and loss rate, while the video score is derived from Video Multi-method Assessment Fusion (VMAF)\cite{NetflixVMAF}. Where $d_{\max}$, $d_{\min}$, $d_{95\mathrm{th}}$, $L$, and $U$ respectively denote the maximum, the minimum, and the 95th percentile queuing delay, packet loss ratio, and bandwidth utilization. 

RBWE, Schaferct, and GCC were deployed in the same controlled environment for online evaluation. We selected approximately 400 emulated test calls, evenly distributed across various bandwidth levels. The Linux Traffic Control (TC) tool was used to precisely emulate bottleneck link capacity and network latency, with a queue length of 50 packets. 
The experimental results are summarized in Table~\ref{tab:online_results}, showing that RBWE achieves the highest overall QoE. Although the QoE evaluation metric differs from the training reward formulation, both emphasize user-perceived quality, and their alignment can be further improved by adjusting the audio-video weighting coefficient $\alpha$. 
RBWE consistently outperforms Schaferct across all evaluation metrics and surpasses GCC in terms of bandwidth utilization and video quality, albeit with a minor trade-off in latency performance. This reflects a fundamental design balance in RTC systems, where bandwidth efficiency and delay sensitivity must be carefully managed.

Figure~\ref{fig:four_images_cdf} presents the CDFs of four key metrics. RBWE consistently outperforms the baseline methods in terms of bandwidth utilization ($S_{rate}$) and video quality ($S_{video}$), achieving a more efficient allocation of network resources and improved user experience. Regarding network score ($S_{network}$), RBWE exhibits intermediate behavior between GCC and Schaferct, striking a balance between adaptability and stability. Although the mean QoE improvement appears marginal, the distribution analysis reveals that RBWE effectively mitigates the tail effect, ensuring more reliable performance across diverse network conditions. Notably, the 10th percentile QoE for RBWE reaches 60.0, marking a 7.0\% increase over Schaferct (56.1) and an 18.6\% increase over GCC (50.6), highlighting its robustness in challenging network scenarios.


Figure~\ref{fig:four_images_trace} provides a detailed comparison of the bandwidth estimation performance across four dynamic network traces. The \textit{true\_capacity} curve represents the actual link bandwidth regulated by TC, while the remaining curves correspond to the bitrate estimates produced by different algorithms. The results highlight that RBWE's superior QoE performance is driven by its ability to accurately track available bandwidth. In contrast, GCC exhibits significant underutilization under rapidly fluctuating conditions, primarily due to the limitations of its AIMD strategy. While Schaferct demonstrates the ability to detect network capacity, its estimates suffer from pronounced fluctuations, leading to instability. Only RBWE achieves an optimal trade-off between responsiveness and stability, effectively adapting to capacity changes while minimizing excessive oscillations.

\section{Conclusion} \label{conclusion}
This paper introduces RBWE, a robust bandwidth estimation framework leveraging offline RL to enhance RTC performance. By integrating Q-ensemble with a Gaussian mixture policy, RBWE effectively captures heterogeneous behavior policies and mitigates OOD risks. Additionally, the proposed fallback mechanism ensures stable and reliable bandwidth estimation during real-world deployment. Experimental evaluations had confirmed RBWE's superiority over heuristic methods and SOTA offline RL algorithm, achieving higher QoE and reducing overestimation errors. Future work will focus on adapting RBWE to meet diverse QoE requirements, ensuring that technological advancements deliver tangible benefits to end-users.

\bibliographystyle{IEEEtran}
\bibliography{ref} 

\begin{thebibliography}{10}
\providecommand{\url}[1]{#1}
\csname url@samestyle\endcsname
\providecommand{\newblock}{\relax}
\providecommand{\bibinfo}[2]{#2}
\providecommand{\BIBentrySTDinterwordspacing}{\spaceskip=0pt\relax}
\providecommand{\BIBentryALTinterwordstretchfactor}{4}
\providecommand{\BIBentryALTinterwordspacing}{\spaceskip=\fontdimen2\font plus
\BIBentryALTinterwordstretchfactor\fontdimen3\font minus \fontdimen4\font\relax}
\providecommand{\BIBforeignlanguage}[2]{{%
\expandafter\ifx\csname l@#1\endcsname\relax
\typeout{** WARNING: IEEEtran.bst: No hyphenation pattern has been}%
\typeout{** loaded for the language `#1'. Using the pattern for}%
\typeout{** the default language instead.}%
\else
\language=\csname l@#1\endcsname
\fi
#2}}
\providecommand{\BIBdecl}{\relax}
\BIBdecl

\bibitem{alvestrand2021rfc}
H.~Alvestrand, ``{RFC 8825: Overview: Real-Time Protocols for Browser-based Applications},'' 2021.

\bibitem{carlucci2016analysis}
G.~Carlucci, L.~De~Cicco, S.~Holmer, and S.~Mascolo, ``{Analysis and Design of the Google Congestion Control for Web Real-Time Communication (WebRTC)},'' in \emph{Proceedings of the 7th International Conference on Multimedia Systems}, 2016, pp. 1--12.

\bibitem{xiao2023ember}
X.~Xiao, M.~Yan, Y.~Zuo, B.~Liu, P.~Ruan, Y.~Cao, and W.~Wang, ``{From Ember to Blaze: Swift Interactive Video Adaptation via Meta-Reinforcement Learning},'' in \emph{IEEE INFOCOM 2023-IEEE Conference on Computer Communications}.\hskip 1em plus 0.5em minus 0.4em\relax IEEE, 2023, pp. 1--10.

\bibitem{zhang2020onrl}
H.~Zhang, A.~Zhou, J.~Lu, R.~Ma, Y.~Hu, C.~Li, X.~Zhang, H.~Ma, and X.~Chen, ``{OnRL: Improving Mobile Video Telephony via Online Reinforcement Learning},'' in \emph{Proceedings of the 26th Annual International Conference on Mobile Computing and Networking}, 2020, pp. 1--14.

\bibitem{eo2022opennetlab}
J.~Eo, Z.~Niu, W.~Cheng, F.~Y. Yan, R.~Gao, J.~Kardhashi, S.~Inglis, M.~Revow, B.-G. Chun, P.~Cheng \emph{et~al.}, ``{OpenNetLab: Open Platform for RL-based Congestion Control for Real-Time Communications},'' in \emph{Proceedings of the 6th Asia-Pacific Workshop on Networking}, 2022, pp. 70--75.

\bibitem{khairy2024acm}
S.~Khairy, G.~Mittag, V.~Gopal, F.~Y. Yan, Z.~Niu, E.~Ameri, S.~Inglis, M.~Golestaneh, and R.~Cutler, ``{ACM MMSys 2024 Bandwidth Estimation in Real Time Communications Challenge},'' in \emph{Proceedings of the 15th ACM Multimedia Systems Conference}, 2024, pp. 339--345.

\bibitem{tan2024accurate}
Q.~Tan, G.~Lv, X.~Fang, J.~Zhang, Z.~Yang, Y.~Jiang, and Q.~Wu, ``{Accurate Bandwidth Prediction for Real-Time Media Streaming with Offline Reinforcement Learning},'' in \emph{Proceedings of the 15th ACM Multimedia Systems Conference}, 2024, pp. 381--387.

\bibitem{lu2024pioneer}
B.~Lu, K.~Wang, J.~Xu, R.~Xie, L.~Song, and W.~Zhang, ``{Pioneer: Offline Reinforcement Learning based Bandwidth Estimation for Real-Time Communication},'' in \emph{Proceedings of the 15th ACM Multimedia Systems Conference}, 2024, pp. 306--312.

\bibitem{prudencio2023survey}
R.~F. Prudencio, M.~R. Maximo, and E.~L. Colombini, ``{A Survey on Offline Reinforcement Learning: Taxonomy, Review, and Open Problems},'' \emph{IEEE Transactions on Neural Networks and Learning Systems}, 2023.

\bibitem{agarwal2025mowglipassivelylearnedrate}
\BIBentryALTinterwordspacing
N.~Agarwal, R.~Pan, F.~Y. Yan, and R.~Netravali, ``{Mowgli: Passively Learned Rate Control for Real-Time Video},'' 2025. [Online]. Available: \url{https://arxiv.org/abs/2410.03339}
\BIBentrySTDinterwordspacing

\bibitem{zhou2019learning}
A.~Zhou, H.~Zhang, G.~Su, L.~Wu, R.~Ma, Z.~Meng, X.~Zhang, X.~Xie, H.~Ma, and X.~Chen, ``{Learning to Coordinate Video Codec with Transport Protocol for Mobile Video Telephony},'' in \emph{The 25th Annual International Conference on Mobile Computing and Networking}, 2019, pp. 1--16.

\bibitem{zhang2021loki}
H.~Zhang, A.~Zhou, Y.~Hu, C.~Li, G.~Wang, X.~Zhang, H.~Ma, L.~Wu, A.~Chen, and C.~Wu, ``{Loki: Improving Long Tail Performance of Learning-based Real-Time Video Adaptation by Fusing Rule-based Models},'' in \emph{Proceedings of the 27th Annual International Conference on Mobile Computing and Networking}, 2021, pp. 775--788.

\bibitem{kostrikov2021offlinereinforcementlearningimplicit}
\BIBentryALTinterwordspacing
I.~Kostrikov, A.~Nair, and S.~Levine, ``{Offline Reinforcement Learning with Implicit Q-Learning},'' 2021. [Online]. Available: \url{https://arxiv.org/abs/2110.06169}
\BIBentrySTDinterwordspacing

\bibitem{an2021uncertainty}
G.~An, S.~Moon, J.-H. Kim, and H.~O. Song, ``{Uncertainty-based Offline Reinforcement Learning with Diversified Q-Ensemble},'' \emph{Advances in neural information processing systems}, vol.~34, pp. 7436--7447, 2021.

\bibitem{royston1982expected}
J.~Royston, ``{Algorithm AS 177: Expected Normal Order Statistics(Exact and Approximate)},'' \emph{Applied Statistics}, vol.~31, no.~2, pp. 161--5, 1982.

\bibitem{NetflixVMAF}
Netflix, ``{VMAF: The Netflix Video Multi-Method Assessment Fusion},'' \url{https://github.com/Netflix/vmaf}, 2016, accessed: 2025-05-02.

\end{thebibliography}

\end{document}